\documentclass[11pt]{article}

\usepackage{graphicx}

\begin{document}

\title{Physics of Relativistic Perfect Fluids}
\author{Bartolom\'{e} C{\sc oll}$^1$ and Joan Josep F{\sc errando$^2$}}
\date{}
\maketitle
\noindent
$^1$ Syst\`emes de r\'ef\'erence relativistes, SYRTE, Obsevatoire de Paris-CNRS, 75005 Paris, France.\\
E-mail: \texttt{bartolome.coll@obspm.fr}\\
\vspace{0cm}\\
$^2$ Departament d'Astronomia i Astrof\'{\i}sica, Universitat
de Val\`encia, 46100 Burjassot, Val\`encia, Spain.\\
E-mail: {\tt joan.ferrando@uv.es}
\begin{abstract}
     A criterion is presented and discussed
to detect when a divergence-free perfect fluid
     energy tensor in the space-time describes an
     evolution in {\em local thermal equilibrium}.

     This
     criterion is applied to the class II Szafron-Szekeres
     perfect fluid space-times solutions, giving a very simple
     characterization of those that describe such thermal evolutions.
     For all of them, the significant thermodynamic
     variables are explicitly obtained.

     Also, the specific condition is given under which the
divergence-free perfect fluid energy tensors may
     be interpreted as an {\em ideal gas}.
\end{abstract}

\section{Physical fluids and energetic evolutions.
 The inverse problem}

        In the absence of nuclear or chemical reactions, and
independent of the external constraints to which it may be
submitted, a material medium is considered here as {\em
physically
characterized} by the specification of its molecular components.

        In a domain $\Omega$ of the space-time, and in the absence
of exterior constraints, every initial configuration  of a
material medium gives rise to a particular evolution. And, to
every one of these evolutions in $\Omega$ an {\em energy tensor}
\cite{1} $ T_f $ corresponds {\em univocally}. Due to the absence
of exterior constraints, the energy tensors so obtained are
divergence-free, $ \nabla \cdot T_f = 0 $.

        Thus, {\it a set} $ {\bf T}_f \equiv \{T_f\} $ {\em of energy
tensors} is associated {\em to every medium} $ f $, namely those
of all its possible unconstrained evolutions in the space-time
domain $\Omega$. We have the diagram of Figure 1.

\begin{figure}
\begin{center}
   \includegraphics[width=12cm]{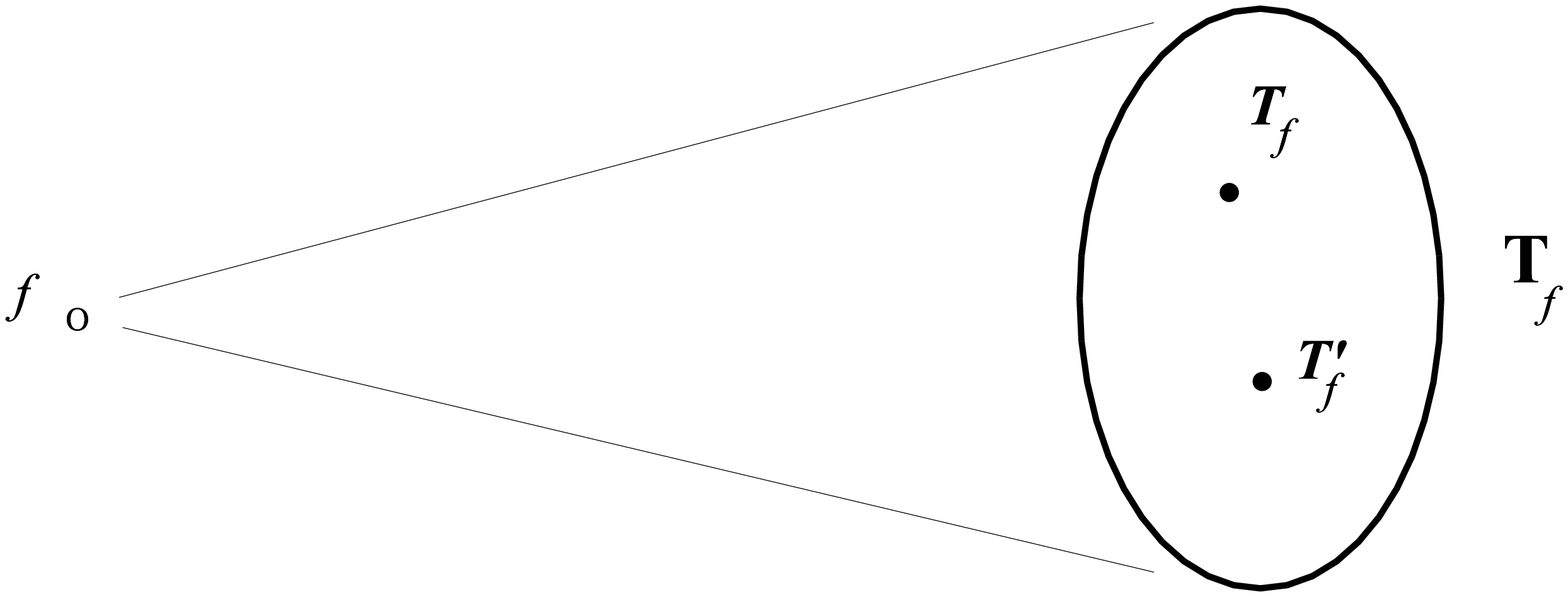}\\
  \caption{}\label{fig1}
\end{center}
\end{figure}

        The energetic description of a particular evolution of the
medium $f$ by the energy tensor $ T_f $, consists, and only
consists, of the specification of the {\em energy density} $ \rho
_f $, the {\em energy flow density} $ q_f $ and the {\em stress
tensor} $ t_f $ of this evolution of $f$ in $\Omega$.

Consequently, the energetic description by $T_f $ {\em  does not
explicitly contain} the physical characterization of the medium,
that is, the specification of its molecular components.

        Thus, a given divergence-free energy tensor $T_f$ could
describe the particular evolution of more than one medium. In
other words, the sets ${\bf T}_f$ and ${\bf T}_{\overline{f}}$ of
all possible unconstrained evolutions of two mediums $f$ and
$\overline{f}$ are not necessarily disconnected as the diagram of
Figure 2 shows.

\begin{figure}
\begin{center}
   \includegraphics[width=10cm]{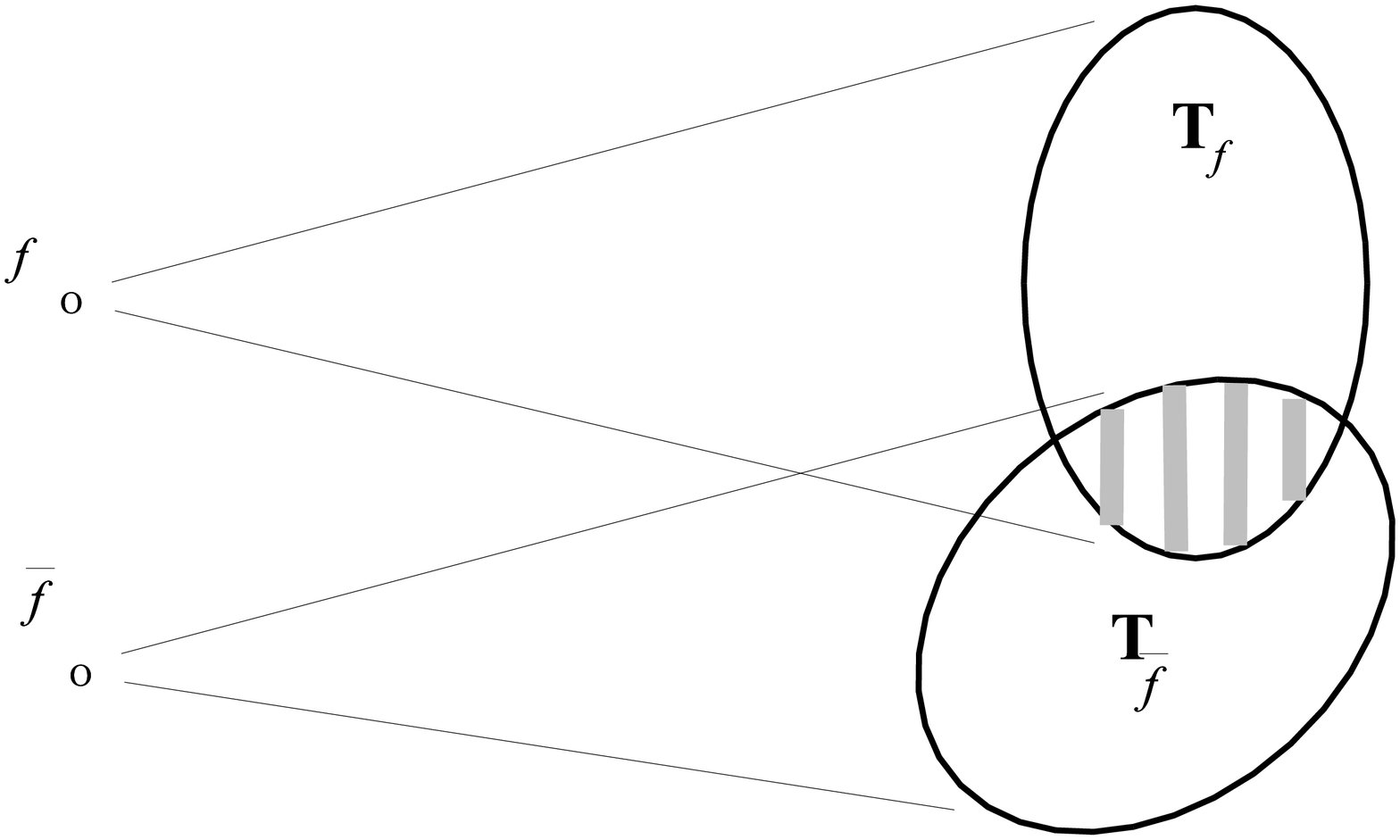}\\
  \caption{}\label{fig2}
\end{center}
\end{figure}

\begin{figure}
\begin{center}
   \includegraphics[width=10cm]{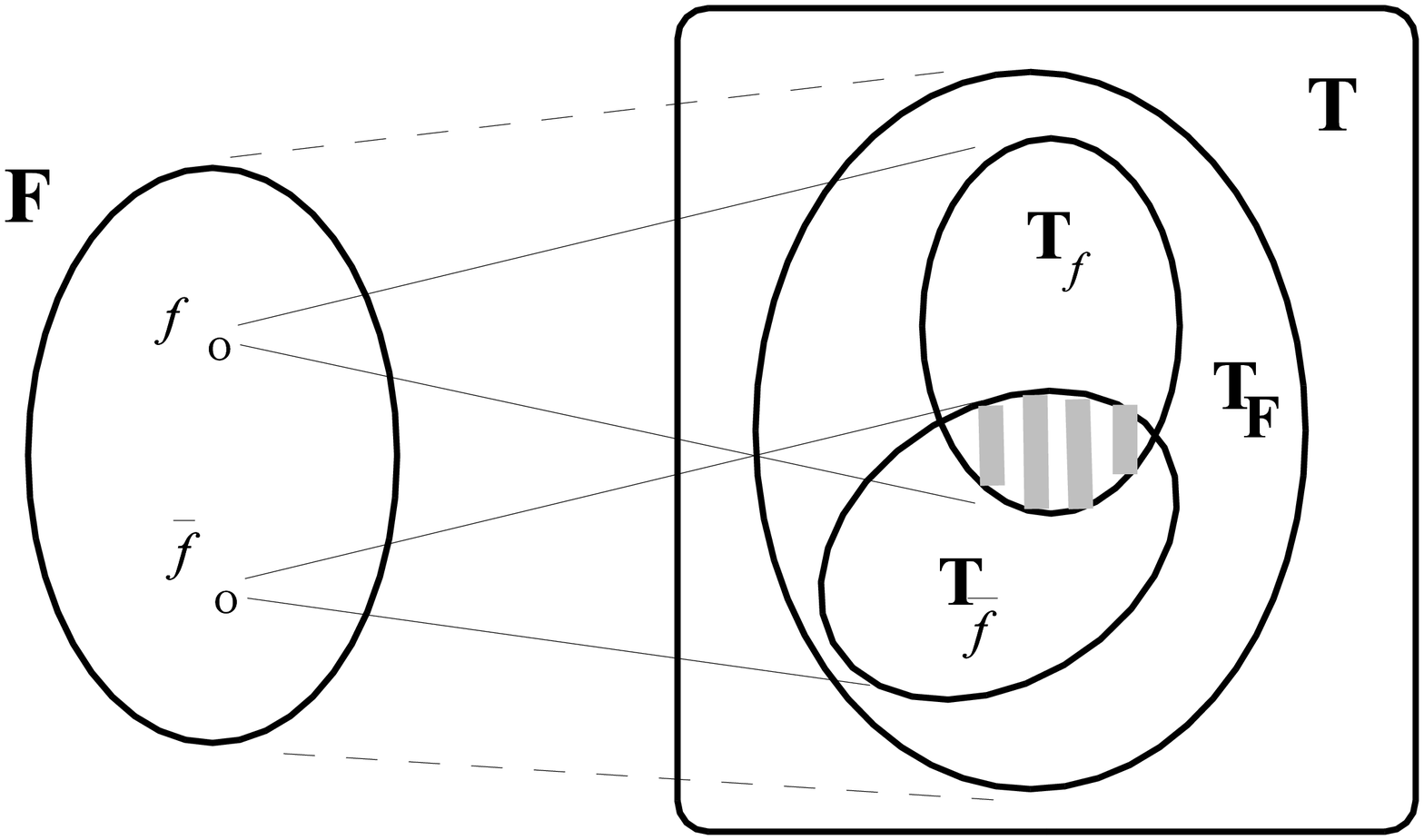}\\
  \caption{}\label{fig3}
\end{center}
\end{figure}

        On the other hand, divergence-free energy tensors $T$ may be
generated theoretically in many ways, which may or may not
correspond to particular evolutions of a physical medium. Thus, if
$\bf T$ denotes the set of all divergence-free energy tensors $T$,
${\bf T} \equiv $ $\{T /  \nabla \cdot T = 0 \}$, $\bf F$ the set
of all mediums $f$, ${\bf F} \equiv $ $\{ f \}$, and ${\bf T}_{\bf
F}$ that of all divergence-free energy tensors $T$ describing all
particular evolutions of all mediums $f$, ${\bf T}_{\bf F} \equiv
$ $\{ T \in {\bf T} / \exists f:  T = T_f , f \in {\bf F} \}$, we
have the diagram of Figure 3.

        Usually, in many (theoretical or experimental) physical
situations one starts from known elements of ${\bf F}$ and looks
for some of the corresponding elements of ${\bf T}$. Here we are
interested in the  {\em inverse problem} that inquires about the
existence
 and characterization of mediums $f$ in ${\bf F} $ corresponding to a given
energy tensor $T$ of ${\bf T}$ (see Figure 4).

\begin{figure}
\begin{center}
   \includegraphics[width=12cm]{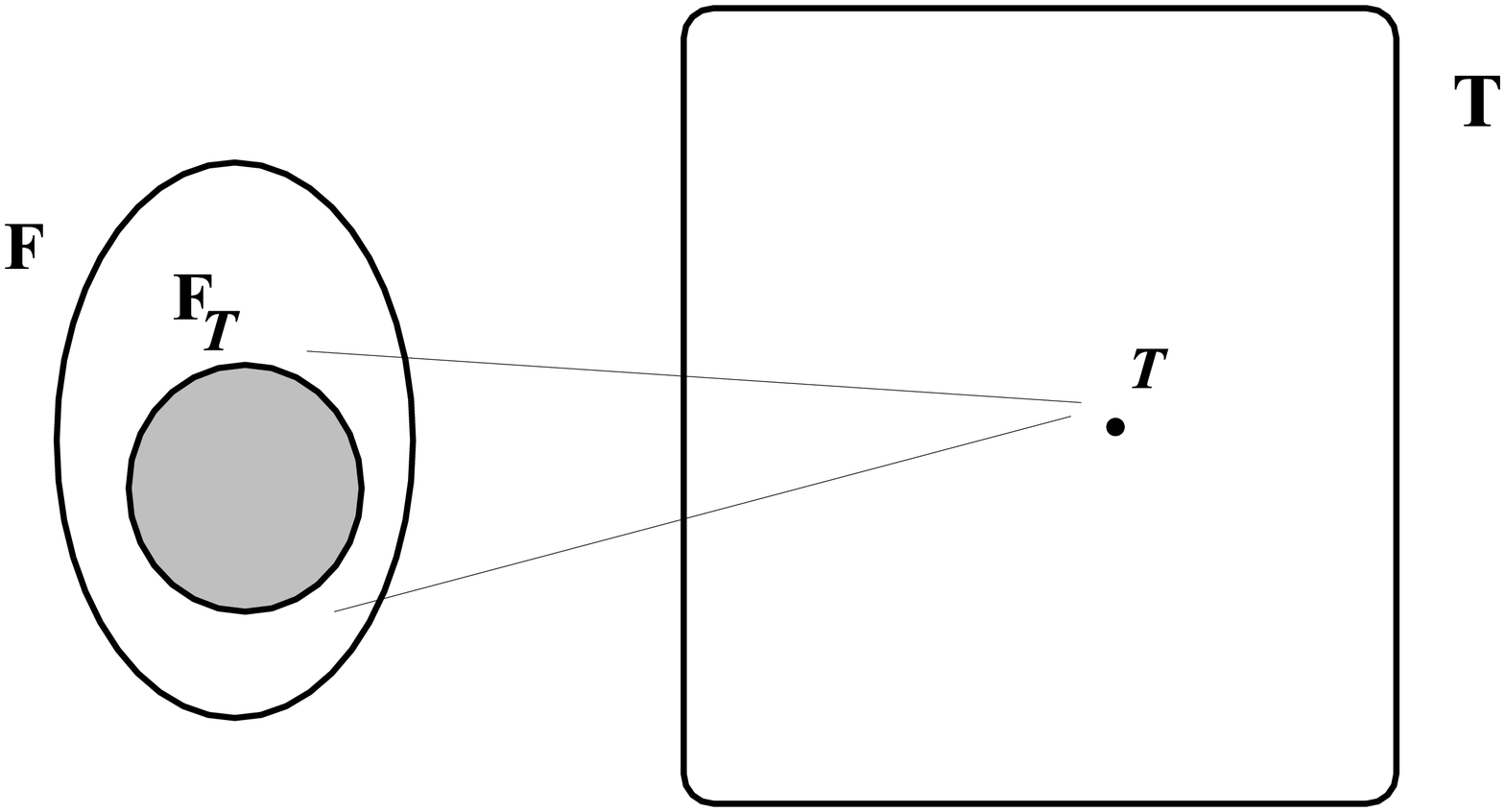}\\
  \caption{}\label{fig4}
\end{center}
\end{figure}

        In this paper we \pagebreak shall restrict the analysis of
  such an inverse problem to the subclass
 $ {\bf T}^* $
 of $ {\bf T} $ constituted by the set of {\em perfect fluid energy tensors},
that is to say, by those divergence-free energy tensors,
\begin{equation}
   \nabla \cdot T = 0  \leftrightarrow  \nabla _{\alpha}T^{\alpha \beta}
= 0 \end{equation}
of the form
\begin{equation}
T = (\rho + p) u\otimes u + p g
\end{equation}

        In spite of its apparent clarity, it is worthwhile asking the
following question: is this inverse problem physically meaningful?

In Section 2 we shall see that additional assumptions are
necessary in order to obtain physically meaningful answers, and the
assumption selected here will be that of {\em local thermal equilibrium}
({\em l.t.e.}).
The necessary and sufficient condition for an energy tensor $T$ of
$ {\bf T}^* $ to be in {\em l.t.e.} will be presented, and some important
properties will
be mentioned.

In Section 3 we shall show how  our condition applies in a particular
example: among the Szekeres-Szafron class II espace-time
solutions, our condition  easily allows us to obtain  {\em all} the
perfect fluid solutions which are in {\em l.t.e.} and  gives the
corresponding matter density, specific entropy and temperature.

Finally, in Section 4, as a further application,  we shall present the
necessary and sufficient conditions for a perfect fluid to be an
{\em ideal gas}.

The proof of all our statements will be given elsewhere \cite{2},
where we will present a deeper analysis of the ideas related to the
"physical meaning" of perfect fluids.

\section{Local thermal equilibrium}

As described above, the set $ T_f $  of energy tensors associated in
$\Omega$ to a fluid $f$ gives the energetic description of
the set of
unconstrained  evolutions that correspond to the different {\em initial
configurations} that the fluid may adopt. As a consequence,  it
is {\em meaningless} to try to find a physical interpretation to a
perfect fluid energy tensor $T$ in $\Omega$ submitted to the sole
\pagebreak
divergence-free condition. The reason is that the system $\{(1),(2)\}$
is not causal, so that not only one energy tensor corresponds to every
initial configuration
, but a partially arbitrary family of such tensors,
which include arbitrarily advanced-retarded-like solutions as well as
Maxwell's demonds-like perturbations of the physical evolution \cite{3}.
In order to select a physical (causal) evolution, one has to
complete the above system conveniently  \cite{4}.
It is known \cite{5} that
its completion with a {\em l.t.e.} scheme is causal.

Here we shall choose to study the inverse problem for a divergence-free
perfect fluid energy tensor under the hypothesis that its evolution
takes place in {\em l.t.e.}.

 Let us remember that a perfect fluid is in {\em l.t.e.} if and only if
{\bf i)} the intensive variables of the fluid are related by {\em equations
 of state}, {\bf ii)} these equations are compatible with the thermodynamic
 relation $Tds = d\epsilon + pdv$, {\bf iii)} the {\em matter density}
 $r \equiv 1/v$ is conserved,
 $\nabla \cdot (ru) = 0$.

As already stated, the {\em l.t.e.} scheme thus defined is a causal
completion
for the system $\{(1),(2)\}$. But it is not a {\em minimal} causal
completion in the sense that it introduces a clearly superabundant
number, not only of equations, but also of unknowns. Because the system
$\{(1),(2)\}$ consists of four equations for five unknowns, the
following question can be asked: does
an {\em equivalent} formulation of the {\em l.t.e.} scheme exits that
is reduced to the {\em addition} to the system of {\em  a unique
equation} relating the five given unknowns?

   This question is far from being academic. If the answer
were positive, not only would we dispose of an easy deductive algorithm
to detect whether or not a given divergence-free perfect fluid energy
tensor evolves in {\em l.t.e.}, but we would also dispose of an
equational
and conceptual simplified version of {\em l.t.e.}: a version that would
show that, contrary to what is indicated by general opinion and
historic development, the concept of {\em l.t.e.}
does not need a {\em thermodynamic}, but only a {\em dynamic},
purely {\em energetic} background.

In fact, in a very different conceptual situation, we
already answered this question positively some years ago
\cite{6}.
 The result is the following theorem:
\vspace{0.3cm}\\
{\bf Theorem} (Coll-Ferrando, 1989): {\em A divergence-free perfect fluid
energy tensor evolves in l.t.e.} if, and only if, {\em the space-time
function}
\begin{equation}
 \chi \equiv \frac{\dot{p}}{\dot{\rho}}
 \end{equation}
{\em called the} indicatrix {\em of l.t.e.},
 {\em depends only on the variables} $p$ {\em and} $\rho$ {\em :}
\begin{equation}
 d\chi \wedge dp \wedge d\rho = 0
 \end{equation}

We believe that this simple result is  very interesting  from the
physical point of view because of the following remarks:

{\bf i)}  This theorem says, in other words, that a relation of the form
\begin{equation}
\dot{p} = \chi (p,\rho )\dot{\rho}
\end{equation}
is a {\em minimal causal completion} to the equations $\nabla .T = 0$ for
a perfect fluid.

{\bf ii)}  Condition (4), $ d\chi \wedge dp \wedge d\rho = 0 $, is a
{\em deductive}  condition of {\em  l.t.e.} for the {\em
hydrodynamic} variables $u$, $p$ and $\rho$.

{\bf iii)} The "iff" character of the theorem implies that this
condition  constitutes an {\em alternative} definition of {\em  l.t.e.}.
And surprisingly enough at first glance, this alternative definition involves {\em only
 hydrodynamic, energetic}
and {\em evolutive} concepts, but {\em not thermodynamic} ones.

{\bf iv)} If the conditions of the theorem are verified, that is, if
the perfect fluid evolves in {\em  l.t.e.}, then the indicatrix $\chi$
becomes a function of state, $\chi (\rho ,p)$, and it
physically represents the square of the {\em velocity of the sound} in
the fluid \cite{7},
\begin{equation}
\chi (\rho ,p) \equiv  v^2_{sound}
\end{equation}

{\bf v)} From the above interpretation, one has the following
necessary condition of {\em physical reality}:
\begin{equation} 0 \leq \chi \leq 1
\end{equation}

In the next Section we have selected a general (family
of) perfect
fluid space-time(s) to show how our theorem is applied in a particular
case.

\section{Class II Szekeres-Szafron space-times}

The Szekeres-Szafron solutions to Einstein Equations constitute one of
the largest families of perfect fluid exact solutions with no
generic symmetries. A class of them,  the so called {\em class II}, is
given by a line element $ds^2$ of the form \cite{8}
\begin{equation}
ds^2 = dt^2 - R^2\left\{ (B + P)^2dx^2 + {dy^2 + dz^2 \over
\left[ 1 + {k \over 4}(y^2 + z^2) \right]^2}\right\}
\end{equation}
with
\begin{equation}
P \equiv {{1\over 2}(y^2 + z^2)U + y V_1 + zV_2  \over
1 + {k\over 4}(y^2 + z^2)}
\end{equation}
where $R(t)$, $U(x)$, $V_1(x)$, $V_2(x)$, $V(x)$ are arbitrary functions,
$k \in \{ 0, \pm 1\}$, and $B(x,t)$ is a solution of the equation:
\begin{equation}
\ddot B + 3{\dot R \over R} \dot B - {kB + U \over R^2} = 0
\end{equation}

In these coordinates the perfect fluid is comoving, and its
pression $p$ and energy density $\rho$ are given by
\begin{equation}
p = - {2R\ddot R + {\dot R}^2 + k \over R^2}
\end{equation}
\begin{equation}
\rho = {2R\dot R\dot B + 3(B + P){\dot R}^2 + k(B + 3P) - 2U \over
R^2(B + P)}
\end{equation}

In order to characterize, among these fluids, all those that evolve in
{\em  l.t.e.}, it is convenient to first detect the structure of
the above
function $B(\rho , p)$. Such a structure is given by the following
lemma:
\vspace{0.3cm}\\
{\bf Lemma 1:} {\em The} general solution {\em of the equation}
\begin{equation}
\ddot B + 3{\dot R \over R}\dot B - {kB+U \over R^2} = 0
\end{equation}
{\em is of the form}
\begin{equation}
B = \alpha a + (k\beta + 1)b + \beta U
\end{equation}
{\em where} $a(x)$ {\em and} $b(x)$ {\em are arbitrary functions and}
$\alpha (t)$ {\em and} $\beta (t)$ {\em verify}
\begin{equation}
\ddot \alpha = -3 {\dot R\over R}\dot \alpha + {k\over R^2}\alpha
\end{equation}
\begin{equation}
\ddot \beta = -3 {\dot R\over R}\dot \beta + {1\over R^2}(k\beta +1)
\end{equation}
\vspace{0.2cm}

The isobaric case $\dot{p}=0$, which according to expression (11)
corresponds to particular choices of $R(t)$, obviously verifies our
condition of {\em l.t.e.}.


In order to select all other
solutions that evolve in  {\em l.t.e.}, it is now easy,
 because $p = p(t)$,  to develop expression (4) directly,
and thus allows us to neglect
time derivatives of $\rho $ and $\chi $. The result is
\vspace{0.5cm}\\
{\bf Theorem 1:} {\em The non isobaric class II Szekeres-Szafron
perfect
fluids that evolve in l.t.e. are those for which}
\begin{equation}
a(x) = 0 \ \ \ \ \ \ \ \ \ \ \ \ \ \ \ \ {\it or} \ \ \ \ \ \ \ \
\ \ \ \ \ \ \ \ a(x) \not= 0 \ , \ \ \ k=0 \ ,\ \ \ U(x)=0
\end{equation}
\vspace{0.1cm}\\

For such fluids it is then easy, starting from the "old" definition of
{\em l.t.e.},  to find their thermodynamic variables. One obtains:
\vspace{0.5cm}\\
{\bf Theorem 2:} {\em For the non isobaric class II Szekeres-Szafron
perfect fluids that evolve in l.t.e., the} matter density $r$
{\em is given by}
\begin{equation}
r = {\phi (u)\over R^3(\beta u+ 1)} \ , \ \ \ \ u \equiv {kb+U \over
b+P}
\end{equation}
{\em for the case} $a(x)=0$, {\em and by}
\begin{equation}
r = {\phi (u)\over R^3(\alpha + u)} \ , \ \ \ \ u \equiv {b+P \over a}
\end{equation}
{\em for the case} $a(x) \not= 0$, $k=0$, $U(x)=0$. {\em The}
specific entropy $s$ {\em and the} temperature $T$ {\em are,
respectively}
\begin{equation}
s = s(u) \ , \ \ \ \ \ \ \ \ \ \ \ T
= {1 \over s'} \left( {\rho + p \over r} \right)'
\end{equation}
{\em where the prime denotes derivative with respect to $u$}.
\vspace{0.4cm}\\

We must note the relatively easy way, and the simple and compact
{\em general} results that may be obtained by direct application of our
above theorem on {\em l.t.e.}. It is worthwhile comparing them with the
partial expressions  obtained, only for
a partial subclass of the class II Szekeres-Szafron solutions,
with the method
claimed by Quevedo and Sussman in reference 8.

\section{When is a perfect fluid an ideal gas?}

Results of the above type allow us to easily find
 {\em all} the families of perfect fluid solutions
 of a general class
that evolve in {\em l.t.e.}. But, what does {\em
a family of perfect fluid solutions} mean physically? Is it {\em a
family of physical fluids},
every one of them in a particular evolution state?, is it {\em a family
of evolution states} of {\em one} particular physical fluid?, or is it a
hybrid family?

There are no known results that allow us to answer this question.
Moreover, even
for a {\em unic} given perfect fluid solution one does not dispose of
simple criteria to detect its plausible physical meaning.

Here, to show how one can solve these problems, we present a
simple criterion for a perfect fluid to be a conventional ideal
gas.
\vspace{0.5cm}\\
{\bf Theorem 3:} {\em The necessary and sufficient condition for a
divergence-free perfect fluid energy tensor, $T = (\rho + p)
u\otimes u + p g$, to represent the l.t.e. evolution of a
classical ideal gas,
\begin{equation}
pv = kT   , \ \ \ \ \   k \equiv {nR\over m},
\end{equation}
with specific internal energy
\begin{equation}
\epsilon = c_v T   ,
\end{equation}
is that the indicatrix $\chi $ be of the form
\begin{equation}
\chi = {\gamma p\over \rho + p}
\end{equation}
with
\begin{equation}
1 \leq \gamma \leq 2 .
\end{equation}
Then, the constants $k$ and $c_v$ are related by}
\begin{equation}
\gamma = 1 + {k \over c_v}
\end{equation}
\vspace{0.2cm}

It is important to note the following points:

{\bf i)} In practice, it is convenient to verify directly that one has
\begin{equation}
d {(\rho + p)\dot p \over p\dot \rho } = 0
\end{equation}

{\bf ii)} If it is the case, then the constant
\begin{equation}
\gamma \equiv  {(\rho + p)\dot p \over p\dot \rho }
\end{equation}
fixes, but only fixes, the quotient $k/c_v$ of the gas. Thus, the values
$\gamma = 5/3$ and $\gamma = 7/5 $ correspond respectively to {\em
monoatomic} and {\em diatomic} ideal gases.

Other gases, not necessarilly ideal ones, will be characterized
elsewhere \cite{2}.

\vfill
\end{document}